\documentclass[amsmath,amssymb, aps, prx,twocolumn,superscriptaddress]{revtex4-2}

\usepackage{hyperref}

\usepackage{braket}
\usepackage{dsfont}
\usepackage{graphicx}
\graphicspath{{images/}}
\usepackage{dcolumn}
\usepackage{bm}
\usepackage{mathtools}
\usepackage[thinc]{esdiff}
\usepackage{textcomp}
\usepackage{subfigure}
\usepackage[dvipsnames]{xcolor}
\usepackage{appendix}
\usepackage{tcolorbox}
\usepackage{multirow}
\usepackage{tikz}

\newtheorem{definition}{Definition}

\usepackage[normalem]{ulem}

\usepackage{orcidlink}

\begin{document}

\title{Graph Coloring via Quantum Optimization on a Rydberg-Qudit Atom Array}
\date{Submitted: \today{}}
\author{Toonyawat Angkhanawin\, \orcidlink{0000-0002-9956-4257}}
\affiliation{\normalfont \textit{Joint Quantum Centre (Durham-Newcastle), Department of Physics, Durham University, South Road, Durham, DH1 3LE, United Kingdom}}
\author{Aydin Deger\,\orcidlink{0000-0002-6351-4768}}
\affiliation{Department of Physics, Clarendon Laboratory, University of Oxford, Parks Road, Oxford OX1 3PU, United Kingdom}
\author{Jonathan D. Pritchard\,\orcidlink{0000-0003-2172-7340}}
\affiliation{Department of Physics and SUPA, University of Strathclyde, Glasgow G4 0NG, United Kingdom}
\author{C. Stuart Adams\,\orcidlink{0000-0001-5602-2741}}
\affiliation{\normalfont \textit{Joint Quantum Centre (Durham-Newcastle), Department of Physics, Durham University, South Road, Durham, DH1 3LE, United Kingdom}}

\begin{abstract}
Neutral atom arrays have emerged as a versatile candidate for the embedding of hard classical optimization problems. Prior work has focused on mapping problems onto finding the maximum independent set of weighted or unweighted unit disk graphs. In this paper we introduce a new approach to solving natively-embedded vertex graph coloring problems by performing coherent annealing with Rydberg-qudit atoms, where different same-parity Rydberg levels represent a distinct label or color. We demonstrate the ability to robustly find optimal graph colorings for chromatic numbers up to the number of distinct Rydberg states used, in our case $k=3$. We analyze the impact of both the long-range potential tails and residual inter-state interactions, proposing encoding strategies that suppress errors in the resulting ground states. We discuss the experimental feasibility of this approach and propose extensions to solve higher chromatic number problems, providing a route towards direct solution of a wide range of real-world integer optimization problems using near-term neutral atom hardware.
\end{abstract}

\maketitle

\section{Introduction}

Many real-world problems in industry and finance can be cast as combinatorial optimization problems \cite{Wurtz2024IndustryProblems}. Whilst some of these lie in the class of \emph{easy} (P) problems that can be solved efficiently in polynomial time using classical hardware, many exist in the class of \emph{hard} (NP) problems that cannot be solved optimally without an exponential growth of the evaluation time, even when exploiting heuristic algorithms offering polynomial-time approximations. However, such problems could be solved optimally with a polynomial growth of evaluation time in non-deterministic machines \cite{Garey1979ComputersSciences,Das2008Colloquium:Computation}. However, despite decades of research in quantum and computer science, it remains an open question whether such non-deterministic machines could be implemented using quantum hardware.

Research into the application of quantum optimization to solving relevant graph problems has explored applications to the Maximum Independent Set (MIS) problem, which consists in finding the largest independent subset of vertices in a graph such that none of the selected vertices are connected by an edge. In the case where each vertex is assigned a weight, this generalizes to the Maximum Weighted Independent Set (MWIS) problem. MIS and MWIS are proven to be NP-complete for both planar graphs \cite{Garey1977TheNP-Complete} and unit disk graphs (UDG) \cite{Clark1990UnitGraphs} with a maximum degree of 3. Recent work has shown that this enables solving underlying MIS and MWIS problems by mapping onto UDG encodings compatible with the native connectivity found in Rydberg atom arrays \cite{Pichler2018QuantumArrays,Pichler2018ComputationalDimensions, Nguyen2023QuantumArrays} and applying routines such as the variational quantum annealing (VQA) \cite{Farhi2000QuantumEvolution,Farhi2001AProblem,Albash2018AdiabaticComputation,Larocca2025} or quantum approximate optimization algorithms (QAOA) \cite{Farhi2014AAlgorithm,Zhou2020QuantumDevices} to obtain solutions.

Neutral atom arrays have emerged as promising platforms for scalable quantum computing \cite{Saffman2010QuantumAtoms,Saffman2016QuantumChallenges,Henriet2020QuantumAtoms,Morgado2021QuantumQubits,Dalyac2024GraphProcessors,Kim2023QuantumGraphs,Adams2019RydbergTechnologies}. By exploiting the strong, long-range interactions of highly excited Rydberg states it is possible to realize a blockade effect that can be leveraged for high-fidelity digital computing \cite{Lukin2001DipoleEnsembles,Levine2019ParallelAtoms,Evered2023High-fidelityComputer,Urban2009ObservationTwoatoms,Gaetan2009ObservationRegime,Jaksch2000FastAtoms}, programmable quantum simulation \cite{Bernien2017ProbingSimulator,Keesling2019QuantumSimulator,Browaeys2020Many-bodyAtoms,Ebadi2021QuantumSimulator,Bluvstein2021ControllingArrays,Scholl2021QuantumAtoms,Asmae2023Universality} or analogue optimization, which is the focus of this paper.

UDGs can be natively embedded into neutral atom arrays by geometrically arranging the atoms, with the edges implemented by placing atoms within a blockade radius of each other. This has resulted in a number of experimental demonstrations of solving both MIS \cite{Ebadi2022QuantumArrays,Dalyac2023ExploringDevices,Byun2022FindingAtoms,kim24} and MWIS \cite{DeOliveira2025DemonstrationShifts,Bombieri2024quantumadiabaticoptimizationrydberg} along with exploration of the requirements for achieving a realistic quantum advantage \cite{Serret2020SolvingBenchmarks,Cain2023QuantumLandscapes,Andrist2023HardnessSpeedups,Schuetz2025quantumcompilationtoolkitrydberg} from these methods. Beyond this, programmable Rydberg-atom graphs with local addressibility can be geometrically arranged to solve other NP-complete problems such as maximum cut (Max-Cut) \cite{Graham2022Multi-qubitComputer,Goswami2024SolvingAnnealers}, integer factorization \cite{Park2023AProblem}, and, especially, 3-satisfiability (3-SAT) \cite{Jeong2023QuantumGraphs} in which the polynomial reduction to the MIS has been proven in \cite{Choi2010AdiabaticProblems}. More generally, these approaches reformulate the problem to that of \emph{quadratic unconstrained binary optimization} (QUBO) \cite{Tan_2021} which can be encoded on atomic arrays using elementary sub-graphs \cite{Byun2024Rydberg-AtomProblems} or gadgets \cite{Nguyen2023QuantumArrays}, with a parity-based approach extending to higher-order constrained binary optimization (HCBO) problems \cite{Lanthaler2023Rydberg-Blockade-BasedOptimization}.

However, many real-world optimization problems involve integer optimization problems (IP) \cite{Chansombat2018Mixinitger,Sawik2005IntegerManufacturing,Sawik2011SchedulingProgramming,Zeng2016ScheduleOptimization} where the decision variables are integers. Given the current development of quantum hardware, there is still no prototype of any physical quantum system into which IPs can be directly encoded. In this work, we focus on solving the \emph{minimum vertex graph coloring problem} (MVGCP) \cite{Formanowicz2012AApplications}, consisting of finding a solution to coloring vertices in a graph such that vertices that share an edge are assigned different colors whilst ensuring the minimum number of colors are used.

Coloring of graphs requiring 3 or more colors is NP-hard, meaning MVGCPs are challenging. These problems arise in a variety of industry applications \cite{Wurtz2024IndustryProblems}, for instance scheduling optimization \cite{Ishihara2020OptimizingControl,Sawik2011SchedulingProgramming} or portfolio selection \cite{Cornuejols2018OptimizationEdition}. Directly solving MVGCP on quantum hardware via formulation as a QUBO is resource intensive, with a graph of $N$ vertices and $k$ colors requiring $\mathcal{O}({kN})$ physical qubits \cite{Kwok2020GraphAnnealing,Goswami2024IntegerAtom}. This has driven development of hybrid quantum-classical approaches seeking to solve MVGCP by using heuristic classical solvers combined with quantum hardware to sample the MIS solution requiring only $\mathcal{O}({N})$ physical qubits \cite{Silva2020MappingAnnealing,Kwok2020GraphAnnealing,Vercellino2023BBQ-mIS:Problems,DaSilvaCoelho2023QuantumProblems}

In this paper, we present a route to natively embed unit-disk graph MVGCP onto a neutral atom platform by performing coherent annealing with Rydberg atom qudits. By coupling to $k$ Rydberg levels, we provide access to a Hilbert space of size $\mathcal{O}(k^N)$ and demonstrate the ability to correctly recover the optimal graph colorings with chromatic numbers $\chi(G)\le k$. This represents a first step towards realizing physical quantum hardware onto which the so-called \emph{quadratic unconstrained integer optimization} (QUIO) can be directly encoded without any mapping to the conventional QUBO \cite{Karimi2017PracticalAnnealers}. 

\section{Overview of main results}\label{section:overview_results}

\begin{figure}[t!]
    \centering
    \includegraphics[width=8cm]{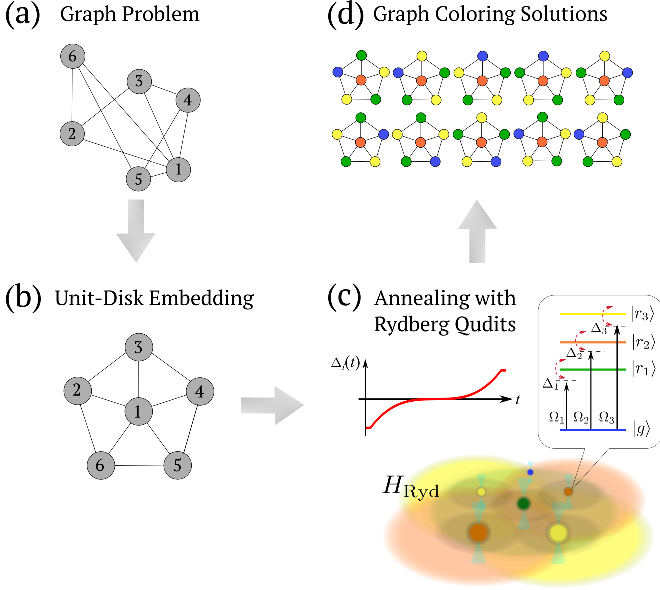}
    \caption{\textbf{Quantum optimization for graph coloring problems.} (a) The original (planar) graph problem is embedded into a corresponding unit-disk graph with compact structure as shown in (b). Here, the vertices and edges of the graph are represented by atoms and their nearest-neighbor interactions. (c) Adiabatic quantum annealing is performed by instantaneously turning on the qudit driving fields $\Omega_i$, and slowly sweeping the detunings $\Delta_i(t)$ from large negative value to positive value (in the direction of red dashed arrows). Here, $i=\{1,2,..,k\}$ where $k$ is the maximum number of Rydberg states (colors) used. (d) The measurement results at the end of the protocol show a set of degenerate (due to the graph symmetry) optimal solutions to the given MVGCP.}
    \label{fig:optimization-protocol}
\end{figure}

We propose and numerically demonstrate a native embedding of the QUIO formulation of a MVGCP on a $k$-chromatic graph with $N$ vertices using a qudit-based Rydberg system. This consists of $N$ ground state atoms, each with an EM field coupling to $k$ Rydberg states. The ground and Rydberg states represent our distinct colors. By performing quantum annealing algorithm on the system, we can find optimal graph colorings for planar graphs. 

The protocol is illustrated in Fig.~\ref{fig:optimization-protocol}. Firstly (a) an original planar graph is mapped to the corresponding unit-disk graph with \emph{compact} structure by the so-called \emph{vertex-to-atom} mapping method, as shown in Fig.~\ref{fig:optimization-protocol}(b). Here, compact structure means the arrangement with maximized numbers of equidistant edges. Each pair of neighboring atoms (representing adjacent vertex pairs) is arranged with a spatial separated less than the Rydberg blockade radius, indicated by the green, orange and yellow shades as in Fig.~\ref{fig:optimization-protocol}(c). This results in the blockade of a double excitation of the corresponding Rydberg state for any pair of neighboring atoms, i.e. a double excitation of the green Rydberg state is blocked within green shaded region. Similarly, double excitation of the orange (yellow) Rydberg states are blocked within the orange (yellow) shaded region. Simultaneously, we ensure that all the atoms do not fall into the small dark shade of their neighbors, in which the energy spectrum will be affected by undesired negative inter-Rydberg interactions, such that the system's ground state could be altered, making quantum annealing inefficient. The quantum annealing algorithm is performed by driving the quantum dynamics from the initial Hamiltonian with an \emph{easy-to-prepare} ground state, i.e. the product state of the atomic ground state $\ket{gg...}$, to the final Hamiltonian whose ground state encodes the solutions to the given MVGCP. Here, the detuning $\Delta_i$ of each Rydberg state $\ket{r_i}$ is adiabatically tuned, as shown in Fig.~\ref{fig:optimization-protocol}(c), to ensure that the annealing state remains in the instantaneous ground state at all times \cite{Albash2018AdiabaticComputation,Farhi2001AProblem,Hauke2020PerspectivesImplementations}. Our results show that the annealing process prepares the system in the lowest energy state, and that this state encodes the solution to the corresponding MVGCP.
In particular, we obtain a degenerate subset of optimal graph coloring solutions, in which their configurations yield exactly the same energy due to the symmetry of the graph, as depicted in Fig.~\ref{fig:optimization-protocol}(d). With higher order of the graph symmetry, it has been found that the quantum annealing becomes more efficient such that graph coloring solutions are returned with higher fidelity. 

Regarding the feasibility of universal graph encoding, due to the restricted range of lattice spacings allowed by the encoding constraints in Eq.(\ref{eq:blockade-condition}), solving MVGCPs on equidistant planar graphs is found to be very effective in qudit-based Rydberg systems, as the unwanted negative inter-Rydberg interactions become insignificant compared to the positive conventional (intra) Rydberg interactions. 
However, to solve MVGCPs on more general planar graphs in which equidistant structures cannot be arranged in two dimensions (2D), we explore alternative encoding strategies such as exploiting three-dimensional (3D) graph embedding to achieve the same connectivity as the original 2D graph whilst maximizing the spacing between qubits that are not linked by an edge. Here, the influence of negative inter-Rydberg interactions on solving MVGCPs have also been analyzed.

The paper is structured as follows. In Sec.~\ref{mainsection:MVGCP}, the mathematical definition of MVGCPs is introduced along with brief reviews of previous related research on the hybrid quantum-classical and quantum approaches. Next, we address the limitations of current quantum hardware, and how our proposed qudit-based Rydberg system could yield advantages over these limitations. In Sec.~\ref{mainsection:MVGCP_Rydberg-qudit}, we introduce the qudit-based Rydberg Hamiltonian, and show how MVGCPs could be encoded into such a Hamiltonian. Details of the problem encoding onto Rydberg-atom graphs are included here. In Sec.~\ref{mainsection:equidistant-graph}, we demonstrate the annealing results of MVGCPs on several equidistant $3$-chromatic graphs, composed of a different numbers of (equilateral) triangle subgraphs. Subsequently, in Sec.~\ref{section:complete-K4}, we demonstrate the graph coloring on non-equidistant $4$-chromatic graphs to highlight the effect of the negative inter-Rydberg interactions, and show how the graph encoding can be improved by exploiting 3D graph embedding. Finally, in Sec.~\ref{mainsection:outlook}, we summarise the advantages offered by our qudit-based Rydberg systems as an alternative route towards native embedding of integer problems, and also discuss the limitations, experimental feasibility and potential to encode other NP-complete problems on this platform. 

\section{Minimum Vertex Graph coloring (MVGCP)}\label{mainsection:MVGCP}

\subsection{Problem statement}\label{section:MVGCP}

Given an undirected graph $G=(V,E)$, where $V$ is a set of vertices and $E$ is a set of edges, a valid solution to the vertex graph coloring problem involves coloring all vertices such that no pair of edge-connected vertices are assigned the same color. A graph coloring that uses $k$ unique colors is called $k$-coloring with the formal definition

\begin{definition}
\label{def:k-coloring}
{\rm ($k$-coloring)} For an undirected graph $G = (V,E)$, the $k$-coloring is a mapping $f_k: V(G) \to \mathbb{C}_k$ with $f_k(v)\neq f_k(w)$ for all $(v,w) \in E(G)$. Here, $\mathbb{C}_k = \{1,2,...,k\}$ is a set of $k$ colors.
\end{definition}

\begin{figure}[t!]
    \centering
    \includegraphics[width=8.6cm]{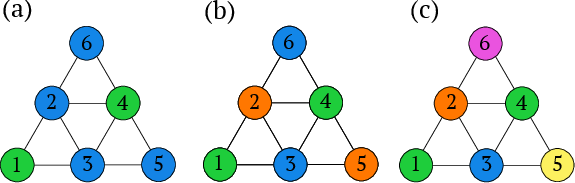}
    \caption{\textbf{Examples of vertex graph coloring.} (a) Invalid graph coloring as adjacent vertices share the same color. (b) An optimal graph coloring with three colors used corresponding to a chromatic number $\chi(G)=3$. (c) Non-optimal graph coloring where all adjacent vertices are assigned with five different colors. This case is referred to as relaxed graph coloring.} 
    \label{fig:graph_coloring}
\end{figure}

In particular, vertex graph coloring is equivalent to partitioning the vertices into $k$ \emph{independent} (stable) sets. The \emph{minimum vertex graph coloring problem} (MVGCP) then consists of finding a valid graph coloring that requires the minimum number of colors. The minimum number $k$ is known as the \emph{chromatic number} $\chi(G)$. Determining the chromatic number of a general graph is widely recognized as NP-hard \cite{Garey1976Thecoloring}, whilst deciding if a graph is colorable with $k$-colors is NP-complete for $k\ge3$ \cite{Karp1972ReducibilityProblems,Garey1976SomeProblems,Garey1977TheNP-Complete,graaf98}. MVGCPs on unit-disk graphs mapped from planar graphs with maximum degree at least 3 are proven NP-complete \cite{Clark1990UnitGraphs,graaf98}. Figure~\ref{fig:graph_coloring} shows example graph colorings, with an invalid solution where vertices with a connected edge share a color in (a) whilst the optimal solution with $k=\chi(G)=3$ shown in (b). However, since in general MVGCPs are known to be hard problems, they are, in practice, relaxed to finding $k$-colorings where $\chi(G) \leq k \leq |V|$ \cite{Titiloye2011QuantumProblem}. This results in sub-optimal graph colorings as shown in (c), and is known as a relaxed coloring which is a valid graph coloring with $k>\chi(G)$.  

\subsection{Classical approaches}
There are a variety of polynomial-time approximate algorithms (PTAAs) which return a non-optimal graph coloring with $k$ no greater than an approximate upper bound relative to the true chromatic number of a problem graph \cite{Pachos2003Polynomial,husfeldt2015graphcolouringalgorithms}. However, due to the NP-hardness of MVGCPs, exact algorithms \cite{Sager1991APP} turn impractical on graphs with hundreds of vertices, hence many heuristic algorithms have become more common in previous research \cite{GALINIER20062547, Johnson1991OptimizationPartitioning}. Among these approaches, heuristic greedy algorithms are widely used such as the Welsh-Powell \cite{Welsh1967AnProblems} or Dsatur algorithms \cite{Brelaz1979NewGraph} which color vertices sequentially, but with different approaches to choosing the vertex ordering based on their degree or saturation degree, respectively. For each graph there exists a perfect vertex order that would return optimal colorings, and the Dsatur algorithm has been proven exact on certain families of graphs such as chordal graphs, cycle graphs and wheel graphs \cite{Yekezare2024OptimalityGraphs}. Another widely used heuristic method is the recursive largest first (RLF) algorithm \cite{Leighton1979AProblems.} which sequentially colors the graph by finding the MIS, assigning these vertices to a given color, and then repeating to find the MIS of the remaining vertices after the previous set is removed.

\subsection{Quantum approaches}\label{section:Quantum_approach}

The simplest approach to solving MVGCP on qubit-based quantum hardware is to cast it as a QUBO of the form \cite{Lucas2014IsingProblems}
\begin{align}\label{eq:qubo}
    H_\mathrm{QUBO} = \sum_{v}(1-\sum_{i=1}^{k}x_{v,i})^2 + \sum_{(u,v) \in E(G)} \sum_{i=1}^{k} x_{u,i}x_{v,i},
\end{align}
where $x_{v,i}$ is a boolean variable representing vertex $v$ with color $i$. This requires $kN$ physical Ising spins for the problem to be embedded in the Ising model, and typically the quadratic constraint term leads to a requirement for all-to-all connectivity of the qubits encoding the boolean variables making this highly challenging for near term quantum hardware. Initial benchmarks of this approach for small problems sizes however showed superior performance for quantum annealing on a D-Wave system compared to simulated annealing \cite{Silva2020MappingAnnealing}.

To mitigate the physical resource and hardware requirements, Fabrikant \emph{et al.} \cite{FabrikantA2002GraphHeuristics} introduced a quantum heuristic method to solve a MVGCP with at most 3 colors, using 2 qubits for encoding each vertex of the graph resulting in an asymptotic performance being polynomial in time. Other work has explored quantum annealing using path-integral Monte Carlo methods \cite{Titiloye2011QuantumProblem}, however this approach is not effective for graphs with large degeneracies. Instead a constrained quantum annealing method has been developed that uses a driving Hamiltonian that encodes constraints without requiring penalty terms, offering a reduction to $N$ physical qubits \cite{Kudo2018Constrainedcoloring}. Tabi \emph{et al} \cite{Tabi2020QuantumEmbedding} implement a space-efficient embedding requiring only $N\text{log}k$ qubits combined with QAOA, however this comes at the cost of deeper circuits which limits performance.

\subsection{Hybrid quantum-classical approaches}\label{section:Hybrid_approach}

Within the development of quantum algorithms there exist several hybrid quantum-classical protocols which seek to off-load part of the classically hard computation onto a qubit-based quantum processor, overcoming the  intensive physical resource requirements for directly mapping MVGCP on a graph with $N$ vertices using $k$ colors into a QUBO acting on $kN$ Ising spins.
Many of these hybrid approaches exploit quantum hardware to iteratively identify the MIS as input for classical heuristic algorithms in a similar approach to RLF. For example, Kwok and Pudenz use MIS solutions to seed a Greedy algorithm \cite{Kwok2020GraphAnnealing}. Vitali \emph{et al.} used a quantum annealer to iteratively solve for \emph{maximal} independent sets (not necessarily the MIS) which are used as a feasible color assignment in a classical branch and bound (BB) method \cite{Morrison2016Branch-and-boundPruning}. Coelho \emph{et al} \cite{DaSilvaCoelho2023QuantumProblems} propose an alternative approach based on the column-generation framework, in which the problem is decomposed into the so-called restricted master problem (RMP) and pricing subproblem (PSP). Here, the RMP is iteratively solved by the classical algorithm with an updated variable (added column) which is a solution to the dual PSP solved with a quantum machine finding the MIS at each step.

\subsection{Qudit-based  approaches}\label{section:qudit_approach}

An alternative approach is to consider algorithms based on using qudits. Wang \emph{et al} \cite{Wang2011Improvedcoloring} introduced a generalization of the Grover algorithm operated on ternary quantum circuits that uses qudits to reduce the complexity of a quantum circuit, resulting in a higher efficiency quantum algorithm. Similar work was carried out by Bravyi \emph{et al} \cite{Bravyi2022Hybridcoloring}, in which the recursive QAOA implemented with hundreds of qutrits has been found to be an efficient algorithm for solving 3-coloring problems in NISQ devices. Recent work from Deller \emph{et al} \cite{Deller2023QuantumSystems} proposes using QAOA with qudit systems to address the electric vehicle charging optimization problem which is mapped onto the MVGCP. This can be extended to formulate a variety of IPs using QAOA, however native qudit based quantum processors have yet to be realized. Amin \emph{et al} realize adiabatic quantum optimization with qudits, in which logical qudits are implemented using many coupled ancilla qubits \cite{Amin2013AdiabaticQudits}, but at the cost of requiring a significant physical qubit overhead. 

In our work, we propose using multi-level Rydberg atoms as a scalable platform for realizing native qudit encodings. To solve MVGCP on a unit disk graph in this case we cast the problem as a spin-glass Potts model \cite{Lucas2014IsingProblems, Inaba2022PottsArchitecture,Wu1982TheModel,Mulet2002coloringGraphs}. To transform from the QUBO representation above in Eq.(\ref{eq:qubo}), we convert the $kN$ binary variables $x_{v,i}$ to $N$ integer variables $n^{(v)}_i$ which encode the color $i$ on vertex $v$, giving rise to the following Potts-like problem Hamiltonian
\begin{align}\label{eq:qubo_P}
    H_{P} \simeq -A \sum_{v \in V(G)}\sum_{i=1}^{k} n^{(v)}_i + B \sum_{(u,v) \in E(G)} \sum_{i=1}^{k} n^{(u)}_i n^{(v)}_i.
\end{align}
In the limit $B\gg A$, the second term prevents vertices connected by an edge from having the same color while the first term maximizes the numbers of repeated colors, and the Hamiltonian therefore encodes a solution to the MVGCP as a ground state. As we will show below, this problem can be directly mapped onto $N$ atoms each with $k$ Rydberg levels.

\section{Graph coloring with Rydberg-atom qudits} \label{mainsection:MVGCP_Rydberg-qudit}

\begin{figure*}[ht!]
    \centering
    \includegraphics[width=17cm]{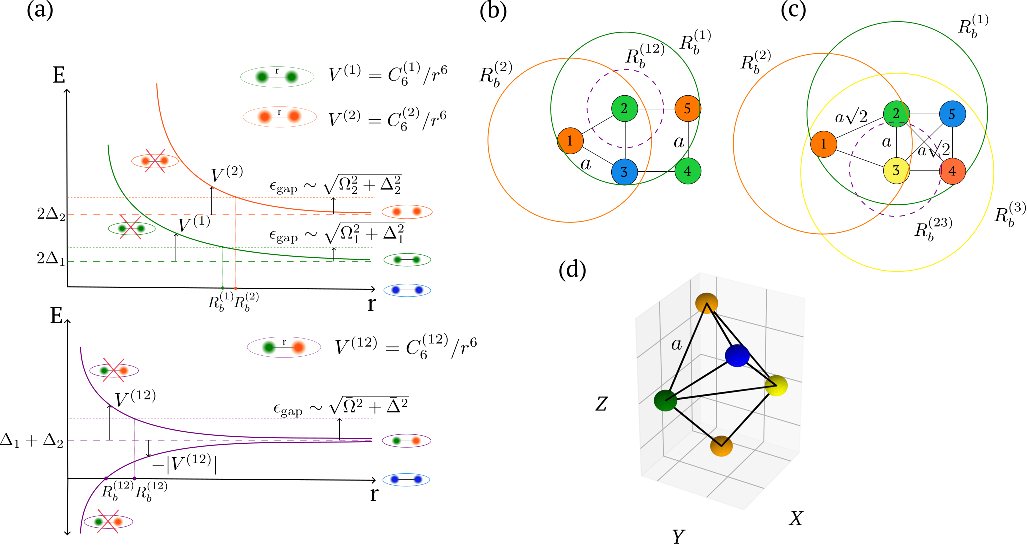}
    \caption{\textbf{Encoding of Rydberg-atom graph.} (a) The energy shifts due to Rydberg interactions between 2 atoms are categorized into two types: 1. (upper) the \emph{intra-Rydberg} interaction denoted by green and orange graphs, respectively, correspond to the double Rydberg excitations $\ket{r_1r_1}$ and $\ket{r_2r_2}$. 2. (lower) the \emph{inter-Rydberg} interaction denoted by the purple curve, correspond to the double Rydberg excitation $\ket{r_1 r_2}$.  Here, the top (and bottom) graph represents the case of $V^{(12)}>0$ (and $V^{(12)}<0$). (b) The equidistant (unit-disk) planar graph implemented by choosing the lattice spacing $a$ such that blockade only connect nearest neighboring (NN) atoms with the conditions: $|V^{(ij)}_{\rm NN}|  \ll \Delta_i \ll V^{(i)}_{\rm NN}$, where $i,j=\{1,2\}$. (c) The non-equidistant (unit-disk) graph implemented by shrinking the lattice spacing, $a$, such that the blockades further connect next-nearest neighboring (NNN) atoms. For NNN atom, the annealing conditions are: $|V^{(ij)}_{\rm NNN}|  \ll \Delta_i \ll V^{(i)}_{\rm NNN}$, and $|V^{(ij)}_{\rm NN}| < \Delta_i$ given the significant increase in $|V^{(ij)}_{\rm NN}|$, see the purple dashed line in the figure. (d) 3D-graph embedding allows the equidistant structure of the non-equidistant graph in (c).}
    \label{fig:graph-encoding}
\end{figure*}

In this section, we provide a detailed description of how the MVGCP problem shown in Fig.~\ref{fig:optimization-protocol} can be solved by mapping onto a qudit-based neutral atom array. Specifically, we consider the case of unit disk graphs, which can be readily realized via the geometric arrangement of atoms using optical tweezers. 

\subsection{Qudit-based Rydberg Hamiltonian}\label{section:Rydberg-Hamiltonian}

As illustrated in Fig.~\ref{fig:optimization-protocol}(c), we consider the case in which an $N$-vertex unit disk graph can be realized using an array of $N$ independent atoms each representing a vertex $v$ of the graph, and edges implemented by placing the relevant vertex atoms adjacent to one another---nearest neighbors (NN).

Each atom consists of a ground state $\ket{g}$ which is coherently coupled to $k$ unique \emph{same-parity} Rydberg states $\ket{r_i}$, where $i=\{1,2,..,k \}$ which encodes our qudit state. Here, use of same parity Rydberg states eliminates flip-flop interactions caused by resonant dipole-dipole interactions, and ensures all interactions can be treated in the van der Waals (vdW) regime with an energy shift $V(R)\propto C_6/R^6$ where $C_6$ is the dispersion coefficient and $R$ is the interatomic separation.

Each Rydberg state is coupled to the ground state using a homogeneous global laser field with Rabi frequency $\Omega_i$ and detuning $\Delta_i$ from state $\ket{r_i}$, resulting in a Hamiltonian of the form
\begin{align}\label{eq:H_Ryd}
H_{\textit{\rm Ryd}} &= \sum_{v \in V(G)}\sum_{i} (\frac{\Omega_i}{2}\sigma^{(v)}_i - \Delta_i n^{(v)}_i )\nonumber \\ 
&+ \sum_{(u,v)\in E(G)} \sum_{i} V^{(i)}(|\boldsymbol{r}_u - \boldsymbol{r}_v|) n^{(u)}_i n^{(v)}_i  \nonumber \\ 
&+ \sum_{(u,v)\in E(G)}\sum_{i<j} V^{(ij)}(|\boldsymbol{r}_u - \boldsymbol{r}_v|) (n^{(u)}_i n^{(v)}_j +n^{(u)}_j n^{(v)}_i) 
\end{align}
where $\sigma^{(v)}_i = \ket{g}_v \bra{r_i} + \ket{r_i}_v \bra{g}$, and $n^{(v)}_i = \ket{r_i}_v\bra{r_i}$ is the projector onto Rydberg state $\ket{r_i}$ of atom labeling vertex $v$.

The first term in the Hamiltonian describes the coherent atom-light interaction due to the laser fields, where the detunings $\Delta_i$ act as rewarding energy for the vertex $i$ to excite to the state $\ket{r_i}$ and the Rabi frequencies $\Omega_i$ add quantum steering. The second and third terms, respectively, represent the \emph{intra-Rydberg}  interactions between pairs of Rydberg atoms in state $\ket{r_i}$ with coefficients $C_6^{(i)}$, and the \emph{inter-Rydberg}  interactions between pairs of atoms in Rydberg states $\ket{r_i}$ and $\ket{r_j}$ with coefficients $C_6^{(ij)}$. They are responsible for penalizing connected vertices that simultaneously excite to the states $\ket{r_ir_i}$ and $\ket{r_i r_j}$, respectively.

The effect of these interaction terms is illustrated in Fig.~\ref{fig:graph-encoding}(a) for the case of two Rydberg levels. In the upper panel we show the pair potential curves for the intra-Rydberg interactions $V^{(1)}(R)$ and $V^{(2)}(R)$ resulting from pairs of atoms in state $\ket{r_1r_1}$ and $\ket{r_2r_2}$. When this interaction exceeds the effective Rabi frequency $\sqrt{\Omega_i^2+\Delta_i^2}$ only a single Rydberg excitation can be created, leading to a blockade for pairs of atoms with a separation below $R_b^{(i)}=(\vert C_6^{(i)}\vert/\sqrt{\Omega_i^2+\Delta_i^2})^{1/6}$. The lower panel shows the inter-Rydberg interaction between atoms in state $\ket{r_1r_2}$. In this case the corresponding blockade condition is satisfied for $R_b^{(ij)}=(\vert C_6^{(ij)}\vert/\sqrt{\bar{\Omega}_{ij}^2+\bar{\Delta}_{ij}^2})^{1/6}$, where $\bar{\Omega}_{ij}=(\Omega_i+\Omega_j)/2$ and $\bar{\Delta}_{ij}=(\Delta_i+\Delta_j)/2$ are the average Rabi frequency and detuning. Note that during annealing, we define the Rydberg blockade radius within the above formulas at the Landau-Zener transition point where $\Delta_i=0$.

\subsection{Encoding of MVGCP on qudit-based Rydberg system}\label{section:Rydberg-encoding}

In the limit $\Omega_i\rightarrow0$ the Rydberg Hamiltonian in Eq.(\ref{eq:H_Ryd}) approximates the Potts-like Hamiltonian of Eq.(\ref{eq:qubo_P}) with $A\rightarrow \Delta_i$ and $B\rightarrow V^{(i)}(\vert\boldsymbol{r}_u - \boldsymbol{r}_v\vert)$ which can be moved inside the summation. Thus by careful choice of parameters we can engineer the ground-state of the interacting Rydberg system to encode the solution of the classical MVGCP problem in a similar manner to the qubit-based Rydberg system being able to solve MIS \cite{Pichler2018QuantumArrays}.

Comparison of the two equations reveals two differences between the classical problem and the Rydberg encoding. The first is the finite-potential tails associated with the vDW interactions, and the second is the additional contribution of the inter-Rydberg couplings. In an ideal system, we would engineer the inter-Rydberg state terms $C_6^{(ij)}=0$, and embed the unit disk graph using edges of length $a<R_b^{(i)}$ for all $i$, such that for $0<\Delta_i \le V^{(i)}(a)$ the ground state matches the MVGCP solution.

However, for real Rydberg states, where in this case we consider the $nS_{1/2}$ Rydberg states of alkali atoms, the inter-Rydberg couplings remain finite and negative with $C_6^{(ij)}<0$, the corresponding interaction tail can be seen as the bottom graph of the lower panel of Fig.~\ref{fig:graph-encoding}(a). Instead, by choosing states with a large separation in principal quantum number $n$, we recover $\vert C_6^{(ij)}\vert < \vert C_6^{(i)}\vert$ for all $i,j$. This introduces additional restrictions on the choice of parameters, such that now, we require the edge spacing $a$ to be chosen such that $R_b^{(ij)}<a<R_b^{(i)}$, and the detunings to be chosen ideally such that $|V^{(ij)}(a)| \ll \Delta_i \ll V^{(i)}(a)$.

A secondary consequence of the negative inter-Rydberg interactions is that the positive energy penalty of an edge-connected coloring violation between a pair of atoms $\ket{r_ir_i}$ can be cancelled out by the negative energy associated with edge coupling to neighboring atoms in state $\ket{r_j}$. To prevent this blockade violation we introduce a lower bound on the detuning to give the constraint
\begin{align}\label{eq:blockade-condition}
 |V^{(ij)}(a)| < \Delta_i < |V^{(i)}(a)+(\alpha-1)V_\mathrm{max}^{(ij)}(a)|~,
\end{align}
where $\alpha$ is the maximum degree of the graph and $V_\mathrm{max}^{(ij)}(a)$ is the largest inter-Rydberg coupling for each $i$.

The procedure above requires tuning parameters such that the intra-Rydberg state blockade radii $R_b^{(i)}$ are comparable. To ensure that the pairwise interactions remain additive and ensure suppression of the unwanted $C_6^{(ij)}$ terms, we use $|n_i-n_j| > 2$ \cite{Cano2012NonadditiveAtoms}. As $C_6\propto n^{11}$, a simple approach to simply re-scale Rabi frequencies such that $(C^{(i)}_6/\Omega_i)^{1/6} \simeq (C^{(j)}_6/\Omega_j)^{1/6}$ quickly becomes unfeasible. Instead we restrict ourselves to the experimentally realizable Rabi frequencies in the range $\Omega/2\pi=1\sim10~$MHz and adjust the final state detuning terms such that $(C^{(i)}_6/\sqrt{\Delta^2_i+\Omega^2_i})^{1/6} \simeq (C^{(j)}_6/\sqrt{\Delta^2_j+\Omega^2_j})^{1/6}$ with $\Delta_i \ge \Omega_i$ and $\Delta_j \ge \Omega_j$.

For equidistant planar graphs these conditions on spacing and relevant interactions can easily be met when performing direct \emph{vertex-to-atom} mapping, with an example of such an MVGCP embedding for a 5-vertex equidistant graph with maximum degree 3 shown in Fig.~\ref{fig:graph-encoding}(b), where the corresponding blockade radii are indicated as colored circles. Here, the interaction distance is adjusted to give only a nearest neighbor (NN) interaction.

For higher degree unit disk graphs, it is possible to embed graphs with up to degree 8 on neutral atom arrays using a blockade radius adjusted to implement next-nearest neighbor couplings as illustrated in Fig.~\ref{fig:graph-encoding}(c). For an all-to-all square, using non-equal separations, vertex 1 can be connected to just two neighboring atoms, where again the minimum spacing is defined by the largest inter-Rydberg blockade length $R_b^{(ij)}$. In this regime, the strong negative interactions become more significant, and care must be taken to adjust parameters carefully to ensure that the condition of Eq.(\ref{eq:blockade-condition}) are met. Alternatively, 3D embeddings can be used as shown in Fig.~\ref{fig:graph-encoding}(d) which implements the same coupling graph as Fig.~\ref{fig:graph-encoding}(c) but with increased spacing between connected vertices to further suppress the unwanted interactions.

\begin{figure}[t!]
    \centering
    \includegraphics[width=7cm]{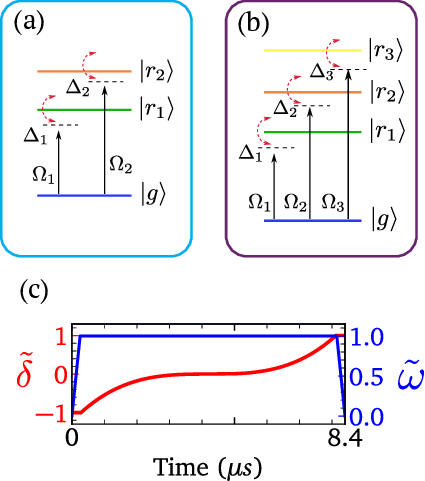}
    \caption{\textbf{Annealing with Rydberg Qudits.} We model annealing for either (a) a ground state plus 2-Rydberg (blue box) or (b) ground state 3-Rydberg (purple box) optimization schemes. (c) The annealing profile shows the normalized time-dependence of the detuning $\tilde{\delta}(t)$ and the Rabi frequencies $\tilde{\omega}(t)$.}
    \label{fig:anneal}
\end{figure}

To investigate the use of Rydberg qudits for performing MVGCP optimization we model two scenarios, using either two or three Rydberg states as shown in Fig.~\ref{fig:anneal}(a) and (b). To meet the requirements above with $\vert C_6^{(ij}\vert<\vert C_6^{(i)}\vert$ we choose the experimentally accessible $nS_{1/2}$ Rydberg states of Rubidium $\ket{r_1}= \vert65S_{1/2}, m_j=1/2 \rangle$, $\ket{r_2} = \vert70S_{1/2}, m_j=1/2 \rangle$ and $\ket{r_3} = \vert75S_{1/2}, m_j=1/2 \rangle$. For these states we extract $C_6$ coefficients by fitting the calculated pair-potentials in the range $R=5-20~\mu$m \cite{Sibalic2017ARC:Atoms}, which for intra-Rydberg interactions gives $\{C_6^{(1)},C_6^{(2)},C_6^{(3)}\} = \{361.0,862.7,1984.5\}~$GHz$\,\mu$m$^6$ and for inter-Rydberg interactions $\{C_6^{(12)},C_6^{(13)},C_6^{(23)}\} = \{-94.1,-35.0,-226.7\}~$GHz$\,\mu$m$^6$.

\subsection{Quantum annealing}\label{section:Quantum-annealing}

To prepare atoms in the ground state of the problem Hamiltonian above, we perform quantum annealing \cite{Pichler2018QuantumArrays,Ebadi2022QuantumArrays,DeOliveira2025DemonstrationShifts} whereby the atoms are initially prepared in state $\ket{g}$ and the global laser fields are adiabatically swept from an initial large negative detuning to a final positive detuning. The parameters are ramped using $\Delta_i(t)=\Delta^{\rm max}_i\tilde{\delta}(t)$ and $\Omega_i(t)=\Omega^{\rm max}_i\tilde{\omega}(t)$, where $\Delta^{\rm max}_i$ and $\Omega^{\rm max}_i$ are chosen to satisfy the encoding constraints above and the normalized time-dependent functions $\tilde{\delta}$, and $\tilde{\omega}$ are defined as \cite{DeOliveira2025DemonstrationShifts,Bernien2017ProbingSimulator}
\begin{align} 
    \tilde{\delta}(t) &= 
\begin{cases} \label{eq:sweeping-protocol-detuning}
    -1,&  0<t<t_i\\
    \frac{8}{\tau^3}(t-t_0)^3,&  t_i<t<t_f\\ 
    1,& t_f<t<T
\end{cases} \\
   \tilde{\omega}(t) &= 
\begin{cases} \label{eq:sweeping-protocol-omega}
    \frac{1}{t_i}t,&  0<t<t_i\\
    1,&  t_i<t<t_f\\ 
    \frac{1}{T-t_f}(T-t),& t_f<t<T
\end{cases}
\end{align}
where $t_0=(t_i+t_f)/2$ and $\tau = t_f-t_i$. The annealing profiles are shown schematically in Fig.~\ref{fig:anneal}(c), where we use $t_i=0.4~\mu$s, $t_f=8~\mu$s, $T=8.4~\mu$s.

To numerically simulate the real-time quantum dynamics, the Trotterization method is used, in which the annealing state is computed by $\ket{\Psi_a(t)} = \prod_{i=0}^{i=p} {\rm e}^{-iH_{\rm Ryd}(t_i)\delta t} \ket{\Psi(0)}$. Here, $\ket{\Psi(0)} = \ket{gg...}$, $\delta t = t_i - t_{i-1}$, $t_0 = 0$, and $t_p=T$, where the annealing time is chosen with $T=8.4 \, \mu s$, and the Trotter time steps $p=300$. Given the error of the method scaling with $\mathcal{O}(T^2/p)$, we Trotterize the time steps by $p = \mathcal{O}(T^2/\epsilon)$ to restrain the error in the admissible scale $\mathcal{O}(\epsilon)$. 

Analysis of the resulting annealing state $\ket{\Psi_a(t)}$ is performed by calculating the dynamical overlap with the ideal MVGC solutions $\ket{\psi}$ using $P_{\psi}(t)= |\langle \psi|\Psi_a(t)\rangle|^2$, or by evaluating the probability of observing a specific computational output solution $\ket{i}$ at the final time $t=T$ as $P_i(T)= |\langle i|\Psi_a(T)\rangle|^2$.

\section{Equidistant (Unit-Disk) Planar Graph} \label{mainsection:equidistant-graph}

\begin{figure*}[t!]
    \centering
    \includegraphics[width=17cm]{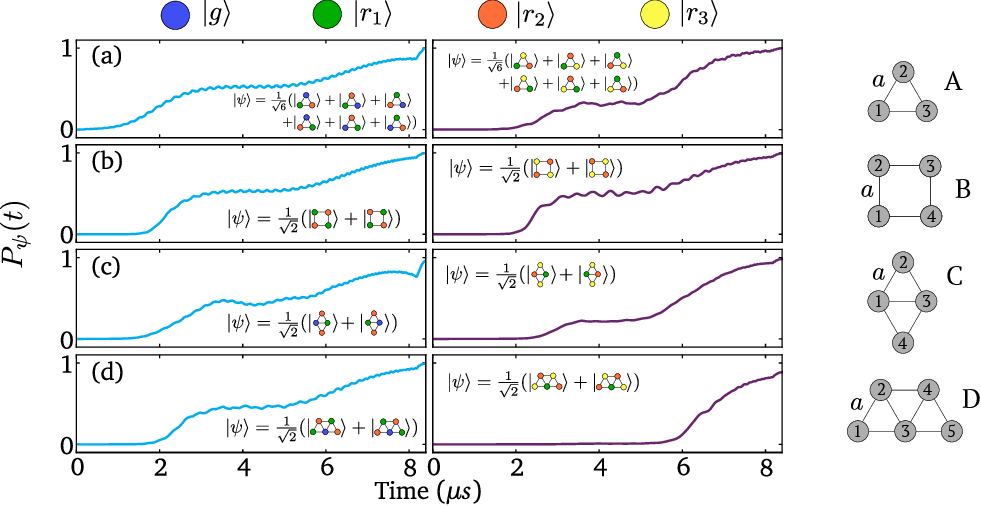}
    \caption{\textbf{Graph coloring for equidistant graphs.} Comparison of 2-Rydberg (left column) and 3-Rydberg (right column) optimizers for equidistant embeddings of (a) a (equilateral) triangle, (b) a square (c) a diamond and (d) a 3-Fan. In all cases, both annealing approaches prepare optimal colorings with high probability approaching 99~\%. For the 3-Fan using the 3-Rydberg annealer, another weak population of optimal colorings with the green and orange inverted are also obtained. Here, 2-Rydberg annealers use $\Delta^{\rm max}_{1,2}/2\pi=8,19~{\rm MHz}$ and $\Omega^{\rm max}_{1,2}/2\pi=3,7~{\rm MHz}$, while 3-Rydberg annealers uses $\Delta^{\rm max}_{1,2,3}/2\pi=5,10,15~{\rm MHz}$ and $\Omega^{\rm max}_{1,2,3}/2\pi=1,2,5~{\rm MHz}$.}
    \label{fig:Planar_graph_and_driving}
\end{figure*}

We first study MVGCPs on planar graphs which can be embedded on neutral atom arrays as equidistant unit-disk graphs as shown in Fig.~\ref{fig:Planar_graph_and_driving}, where all neighboring atoms are spaced with identical distance $a$. For the annealer with two Rydberg states (2-Rydberg optimizer) we use Rabi frequencies $\Omega_{1,2}^\mathrm{max}/2\pi=3,7$~MHz, resulting in blockade radii of $R_b^{(1),(2),(12)}=7.02, 7.05$ and $5.15~\mu$m respectively. These Rabi frequencies are chosen to yield comparable blockade radii. Graphs are then embedded using $R^{(12)}_b < a < 0.8R^{(1)}_b$, leading to lattice spacings approximately in the range $5.15 ~\mu {\rm m} < a < 5.62 ~\mu{\rm m}$. Importantly, to satisfy the encoding constraints in Eq.(\ref{eq:blockade-condition}), the detunings for annealing are chosen as $\Delta^{\rm max}_{1,2}/2\pi = 8,19~\rm{MHz}$. 

For the annealer with three Rydberg states (3-Rydberg optimizer), the chosen Rabi frequencies $\Omega_{1,2,3}^\mathrm{max}/2\pi=1,2,5$~MHz give the corresponding blockade radii $R_b^{(1),(2),(3)}=8.44,8.69,8.55~\mu$m for intra-Rydberg couplings, and $R_b^{(12),(13),(23)}=6.30,4.76,6.32~\mu$m for inter-Rydberg interactions. The lattice spacings of embedded unit disks are then tuned such that $R^{(23)}_b < a < 0.8R^{(1)}_b$, leading to the range around $6.32 ~\mu{\rm m} < a < 6.76 ~\mu{\rm m}$. Again, we choose $\Delta^{\rm max}_{1,2,3}/2\pi = 5,10,15~\rm{MHz}$ to satisfy the encoding constraints. Details on these parameter choices can be seen in Appendix~\ref{Appendix:parameter-choosing}, including the coordinates for all graphs used in the paper given in Table~\ref{table:graph-implementation} along with the actual lengths of lattice spacing in Table~\ref{table:lattice-spacing}.

\subsection{Cycle graphs ($C_N$)} \label{section:cycle-graph}
We begin by considering cycle graphs where every vertex has degree $2$, resulting in a triangle ($C_3$) and square ($C_4$) geometries. Annealing results are shown in Fig.~\ref{fig:Planar_graph_and_driving}(a) and (b), where for both graphs the qudit annealer is able to solve the MVGCP providing an optimal coloring using either 2-Rydberg or 3-Rydberg optimization schemes. For the triangle with chromatic number $\chi(G)=3$, the result is an equal superposition of 6 possible output states due to the underlying $\mathbb{S}_3$ symmetry. 

For the square with $\chi(G)=2$, even for the 3-Rydberg optimizer we recover optimal solutions using $\ket{r_2}$ and $\ket{r_3}$ only, demonstrating that this approach is suitable for efficiently coloring graphs with $\chi(G)<k$. Extending to higher order $C_N$ cycle graphs, we conclude that the chromatic number $\chi(G)=2$ with even $N$, and $\chi(G)=3$ with odd $N$. However, these graphs are not NP-complete, since their every vertex only hold degree 2 \cite{Garey1976SomeProblems,Clark1990UnitGraphs}. 

\subsection{Graphs with maximum degree $\ge3$} \label{section:equidistant-max-degree3}
To demonstrate the qudit optimization in a non-trivial regime we perform MVGCPs on equidistant planar graphs with maximum degree 3 and 4, for example the Diamond (C) and 3-Fan (D) graphs in Fig.~\ref{fig:Planar_graph_and_driving}(c) and (d), respectively. As before, we demonstrate the ability to find optimal $\chi(G)=3$ coloring solutions with high probability for these graphs using either the 2-Rydberg or 3-Rydberg optimization schemes, where for the 2-Rydberg optimizer the ground state provides a label for the third color. We note that the ground state is not expected to be a valid color for all graphs as will be discussed below.

\begin{figure*}[ht!]
    \centering
    \includegraphics[width=17cm]{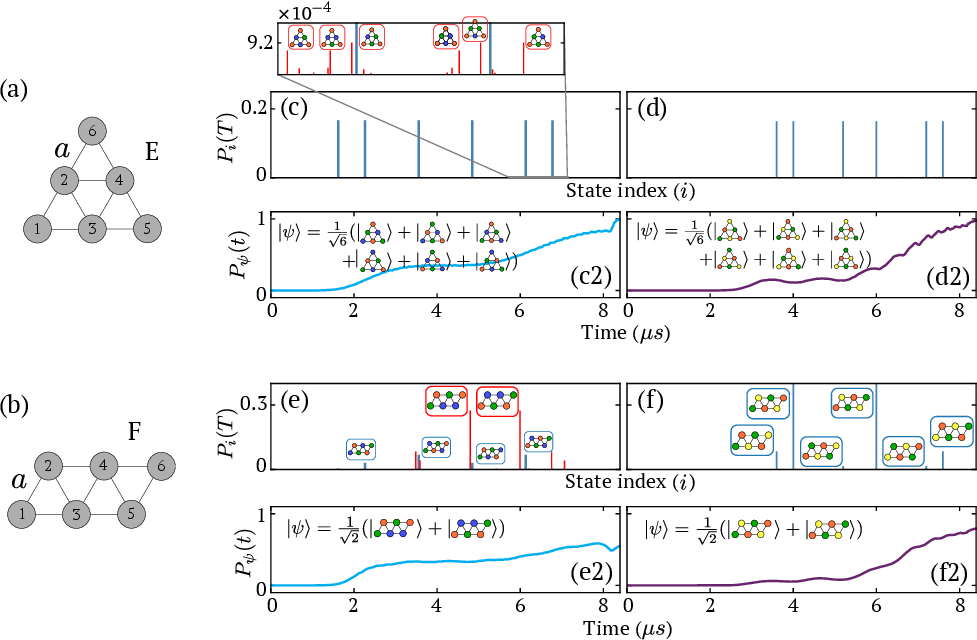}
    \caption{\textbf{Triangular lattice with 6 vertices.} Graph colorings (GCs) on (a) triangle- and (b) ladder-shaped triangular lattices. For the triangle-shape, the 2- and 3-Rydberg annealers yield an equal superposition of 6 degenerate $\mathbb{S}_3$ optimal colorings with fidelity up to $99~\%$ and $97~\%$ as illustrated in (c2) and (d2), respectively. The solutions are in agreement with the final state decompositions shown in (c) and (d). In the ladder-shape, the solutions containing GC defects returned by the 2-Rydberg optimizer are denoted by the red lines in (e). Instead, by employing the 3-Rydberg optimizer in (f) the GC defects are removed, and 3 sets of different $\mathbb{Z}_2$ optimal colorings with unequal portions are observed. The corresponding annealing dynamics are computed in (e2) and (f2). Here, the 2-Rydberg optimizer uses $\Delta^{\rm max}_{1,2}/2\pi=12,14~{\rm MHz}$ and $\Omega^{\rm max}_{1,2}/2\pi=3,7~{\rm MHz}$, and the 3-Rydberg optimizer uses $\Delta^{\rm max}_{1,2,3}/2\pi=5,10,15~{\rm MHz}$ and $\Omega^{\rm max}_{1,2,3}/2\pi=1,2,5~{\rm MHz}$.}
    \label{fig:S3_results}
\end{figure*}

Next, we extend the 3-Fan graph to the case of four equilateral triangle subgraphs in two possible configurations. In the first case, the $4^{\rm th}$ subgraph is added from the above of the 3-Fan graph, forming a \emph{triangle-shape} triangular lattice shown as graph E in Fig.~\ref{fig:S3_results}(a). In the second example, the $4^{\rm th}$ subgraph is added from the right-hand side of the 3-Fan graph, forming a \emph{ladder-shape} triangular lattice shown as graph F in Fig.~\ref{fig:S3_results}(b). Here, both graphs have maximum degree 4. In this section, we investigate the difference between using the 2- and 3-Rydberg optimizers to solve the MVGCPs on both graphs, since significantly different graph coloring results are obtained in the two cases. Here we use the same annealing parameters as before, with the only change being the 2-Rydberg detuning is $\Delta^{\rm max}_{1,2}/2\pi=12,14~{\rm MHz}$.

\subsubsection{Triangle-shape triangular lattice} \label{section:equidistant-max-degree4}

The triangular lattice (graph E) possesses a higher order $\mathbb{S}_3$ symmetry compared with the Diamond (C) and 3-Fan (D) graphs above which only manifest $\mathbb{Z}_2$ symmetry. For the optimization results in Fig.~\ref{fig:S3_results}(c) and (d) using the 2-Rydberg and 3-Rydberg optimizers, we find optimal 3-chromatic colorings $\chi(G)=3$ with fidelities of $99.1~\%$ and $97.2~\%$ respectively. Due to the graph symmetry, the resulting wavefunction is an equal superposition of 6 degenerate states corresponding to permutations of the Rydberg states as shown in the corresponding annealing dynamics in Fig.~\ref{fig:S3_results}(c2) and (d2).

An interesting observation for this problem is that each set of colors in the optimal solutions is a maximal independent set (mIS), but none corresponds to the maximum independent set (MIS). Hybrid algorithms that sequentially color graphs using MIS would result in a suboptimal solution requiring 4 colors; see the zoomed plot in Fig.~\ref{fig:S3_results}(c) which shows the decomposition of the final annealing state into the possible basis states. Our simulations show that these MIS-like solutions (red bars) are strongly suppressed in our annealing process.

\subsubsection{Ladder-shape triangular lattice}
Next, we consider the ladder-shape triangular lattice illustrated in Fig.~\ref{fig:S3_results}(b), where the symmetry of the graph belongs to the $\mathbb{Z}_2$ symmetry allowed by two $\pi$-rotations of the graph. Here the results of optimization in Fig.~\ref{fig:S3_results}(f) show that the 3-Rydberg optimizer yields the optimal coloring solutions whilst the 2-Rydberg case returns an invalid solution with a pair of $\mathbb{Z}_2$ degenerate states containing neighboring ground-state atoms connected by an edge with around 55~\% fidelity. Analysis of the final state decomposition in Fig.~\ref{fig:S3_results}(e) supports this, with the dominant contributions arising from invalid basis states (red bars) with strong suppression of the valid graph coloring states (blue bars).

This failure of the 2-Rydberg system to correctly prepare optimal graph colorings arises due to the lack of interaction between neighboring atoms in the ground state $\ket{g}$, meaning the system energetically favors the configurations of the invalid graph coloring states (red bars) where any two orange vertices with the strongest repulsive interaction are next-next-nearest neighbor (NNNN) to each other. Therefore, to properly solve the MVGCP on such a graph, one needs to employ the 3-Rydberg optimizer.

Results for annealing with the 3-Rydberg optimizer are shown in Fig.~\ref{fig:S3_results}(f) and (f2). Here, the temporal evolution shows the annealing process prepares a set of optimal coloring solutions comprised of an unequal superposition of the six $\mathbb{Z}_2$ states with a fidelity up to $90\%$ in total. These states have almost equal energies, as their energies only differ by the next-next-nearest (NNNN) and next-next-next-nearest (NNNNN) neighbor inter-Rydberg interactions, i.e. $V^{(ij)}_{\rm NNNN} = C^{(ij)}_6/(2a)^6$ and $V^{(ij)}_{\rm NNNNN} = C^{(ij)}_6/(a\sqrt{7})^6$, which are essentially negligible. According to the energy spectrum in Fig.~\ref{fig:spectrum-analysis}(f3) analyzed in Appendix \ref{Appendix:symmetry}, the energy levels of these three groups of different $\mathbb{Z}_2$ states merge into almost the same energy level. 

These results demonstrate that for robust coloring of any target graph, we require $k\ge\chi(G)$ Rydberg states to ensure coloring solutions are represented by populations of strongly interacting Rydberg states.

\begin{figure*}[t!]
    \centering
    \includegraphics[width=15.5cm]{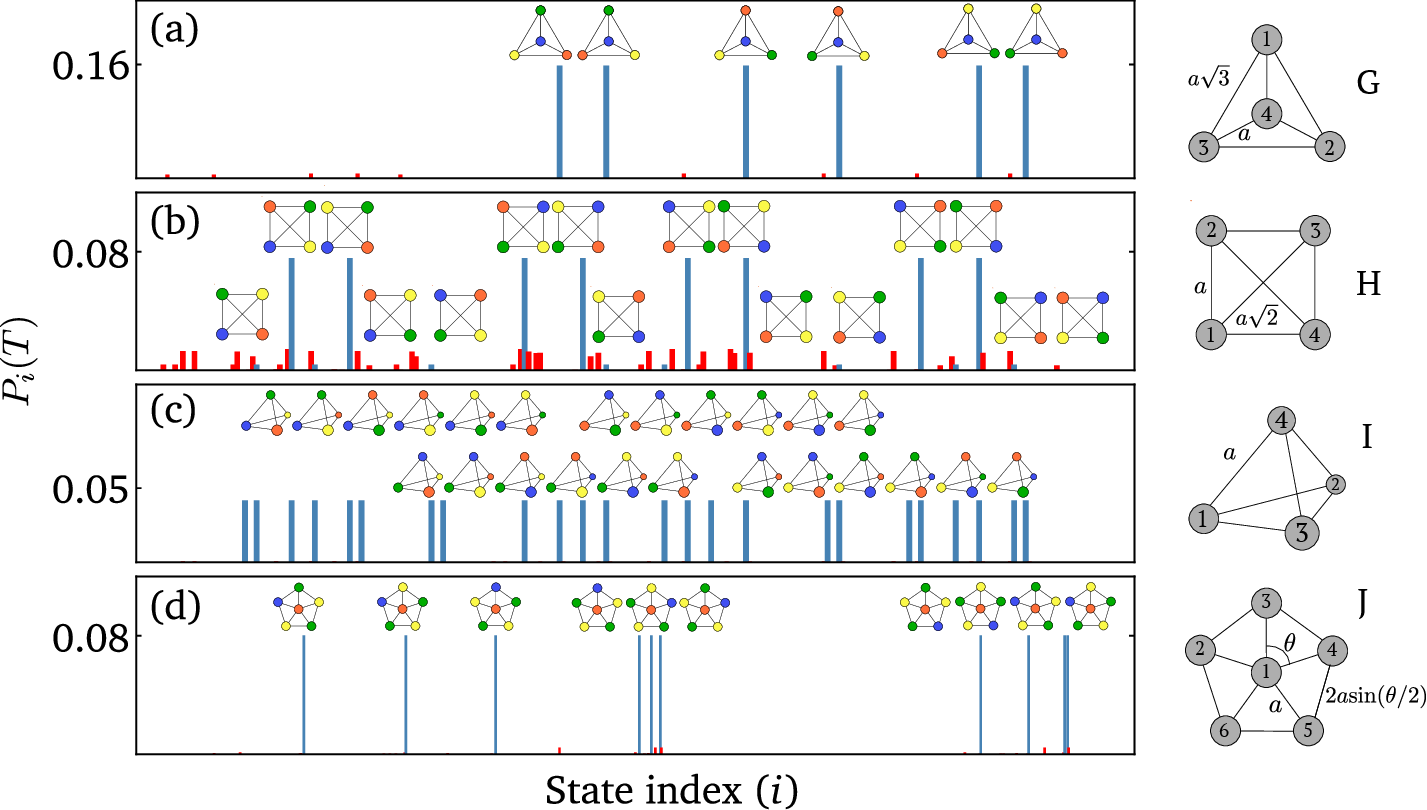}
    \caption{\textbf{Graph coloring on non-equidistant graphs.} The 3-Rydberg optimizer using $\Delta^{\rm max}_{1,2,3}/2\pi=5,10,15~{\rm MHz}$ and $\Omega^{\rm max}_{1,2,3}/2\pi=1,2,5~{\rm MHz}$ is employed to solve the $K_4$ graphs rearranged into three different spatial geometries: (a) the triangle-shape (graph G), (b) the square-shape (graph H) and (c) the equidistant tetrahedron (graph I) in 3D. The Rydberg annealer yields highly-degenerate solutions, arising from the $\mathbb{D}_3$ ($\mathbb{S}_3$), $\mathbb{D}_4$ and $\mathbb{S}_4$ symmetry, respectively. Note that, the energy spectrum of the square-shape (b) is perturbed by significant negative inter-Rydberg interactions, leading to low fidelity  optimal graph colorings (blue) with significant mixings with incorrect colorings (red). This is solved using the 3D embedding shown in (c). (d) The pentagon (graph J) as a wheel graph with six vertices ($W_6$) with a short edge $a$ and long edge $2a~{\rm sin}(\theta/2)$ with $\theta=72^o$. The annealer yields three sets of 10 degenerate states with the total fidelity $95.1\%$. Here, the 3-Rydberg optimizer use $\Delta^{\rm max}_{1,2,3}/2\pi=2.5,10,15~{\rm MHz}$ and $\Omega^{\rm max}_{1,2,3}/2\pi=2,3,5~{\rm MHz}$}
    \label{fig:K4_results}
\end{figure*}

\section{Non-equidistant (unit-disk) graph}\label{mainsection:non-euqidistant-graph}

\subsection{Complete graphs with four vertices ($K_4$)}\label{section:complete-K4}

In this section, we consider MVGCPs on 4-chromatic planar graphs whose corresponding compact unit-disk graphs are non-equidistant when embedded on the atom array. Due to the limitation of computational resources at hand, we restrict to 4-chromatic graphs that can be colored with three distinct Rydberg states, and therefore the 3-Rydberg optimizers with the same sweeping parameters described in Sec.~\ref{mainsection:equidistant-graph} are employed. 

Due to geometry, an equidistant $K_4$ graph is forbidden in 2-dimensional space (2D), but is
allowed in 3-dimensional space (3D). We explore a range of $K_4$ graph symmetries in both 2D and 3D as depicted in Fig.~\ref{fig:K4_results}(right), the coordination of each graph can be seen Tab.~\ref{table:graph-implementation} in Appendix~\ref{Appendix:parameter-choosing}. The different geometries are expected to give rise to different orders of degeneracy in resulting optimal colorings. 
The first spatial arrangement depicted as graph G in Fig.~\ref{fig:K4_results} possesses the $\mathbb{S}_3$ symmetry, hence leading to a set of $3!$ ($6$)-fold degenerate optimal colorings verified by the decomposition of the final annealing state in Fig.~\ref{fig:K4_results}(a). For the second spatial arrangement with the square shape depicted as graph H, a higher order $\mathbb{D}_4$ symmetry is observed, yielding two sets of $8$-fold degenerate optimal colorings including four $\pi/2$-rotations and four mirror reflections, where each is represented by blue bar in Fig.~\ref{fig:K4_results}(b). However, the solutions are presented only with the total fidelity $65.3\%$ as a result of the negative interactions $V^{(ij)}$ that disrupt the energy spectrum of the system. The details of this negativity and their influences are explained in Appendix~\ref{Appendix:negativity-c6}. 

Building on these results, one can infer that identical connectivity of each vertex is desirable. Hence we need to
rearrange the square $K_4$ graph into 3D space, as illustrated in Fig.~\ref{fig:K4_results}(c). In this geometry, the entire graph looks completely identical at every vertex---each vertex is incident to three edges with identical length. The qudit-based Rydberg Hamiltonian Eq.(\ref{eq:H_Ryd}) in this case is able to perfectly simulate the low-energy effective Hamiltonian of the spin-glass Potts model \cite{Wu1982TheModel} with the highest order of symmetry belonging to the $\mathbb{S}_4$ group, the permutation of a group with four elements, giving rise to the degeneracy of order $24$ ($4!$) in expected solutions. These $24$ optimal colorings are obtained in the final annealing  with the total fidelity $98.5\%$, as indicated by 24 blue bars in Fig.~\ref{fig:K4_results}(c). Given the conservation of symmetry along the whole annealing dynamic, the annealing state is energetically confined within the subspace of $\mathbb{S}_4$ symmetry in which the energy spectrum of the twenty-four $\mathbb{S}_4$ degenerate ground states merge into the same energy level as seen in Fig.~\ref{fig:K4_spectrum}(b) in Appendix~\ref{Appendix:negativity-c6}. Despite a vast volume of the $\mathbb{S}_4$ subspace spanned by the 24 degenerate states, their energy spectrum is well separated from the first excited states (red) where there exists a pair of intra-Rydberg double excitations for the `green' Rydberg state. This large energy gap aids the annealing process as it strongly suppresses population of higher excited states that contain sub-optimal or invalid coloring solutions. However, this argument should be properly justified with an additional scaling convergence test, which we will leave for future work. 

\subsection{Pentagon graph ($W_6$)}

Finally, we consider another non-equidistant graph demonstrated in Fig.~\ref{fig:optimization-protocol}, the so-called \emph{wheel} graph with six vertices ($W_6$) as depicted in graph J in Fig.~\ref{fig:K4_results}(d). Since there exists nearest (NN) and next-nearest (NNN) neighbor edges, the negative inter-Rydberg $V^{(ij)}$ interactions become significant, hence we choose another set of three Rydberg states with reduced inter-Rydberg interactions:  $\ket{r_1}= \vert60S_{1/2}, m_j=1/2 \rangle$, $\ket{r_2} = \vert65S_{1/2}, m_j=1/2 \rangle$ and $\ket{r_3} = \vert75S_{1/2}, m_j=1/2 \rangle$, giving the following intra- and inter-Rydberg couplings $\{C_6^{(1)},C_6^{(2)},C_6^{(3)}\} = \{138.9,360.7,1948.4\}~$GHz$\,\mu$m$^6$ and for inter-Rydberg interactions $\{C_6^{(12)},C_6^{(13)},C_6^{(23)}\} = \{-28.5,-8.0,-34.9\}~$GHz$\,\mu$m$^6$. With the Rabi frequencies $\Omega^{\rm max}_{1,2,3}/2\pi=2,3,5~{\rm MHz}$ and the NN (NNN) edge of length $4.10~(4.82) ~\mu{\rm m}$, the resulting Rydberg interactions follow $V^{(12),(13),(23)}_{\rm NN}/2\pi=-6.0,-1.7,-7.3$~MHz, $V^{(12),(13),(23)}_{\rm NNN}/2\pi=-2.3,-0.6,-2.8$~MHz and $V^{(1),(2),(3)}_{\rm NNN}/2\pi=11.0,28.6,154.7$~MHz. In order to satisfy the encoding constraints in Eq.(\ref{eq:blockade-condition}) such that next-nearest neighboring (NNN) atoms, i.e. any pairs of neighboring atoms on the wheel edge, do not stay in the blockade-violated state $\ket{r_1r_1}$, Eq.(\ref{eq:blockade-condition}) becomes $|V^{(12),(13)}_{\rm NNN}| < \Delta_1 < |V^{(1)}_{\rm NNN}+V^{(12)}_{\rm NN}+V^{(13)}_{\rm NNN}|$, i.e. $2.3,0.6 <  \Delta_1/2\pi < 4.4 ~{\rm MHz}$. Likewise, in order for next-nearest neighboring (NNN) atoms to not stay in the blockade-violated state $\ket{r_2r_2}$, the corresponding constraint becomes $2.3,2.8 < \Delta_2/2\pi < 19.1~{\rm MHz}$. Since $V^{(3)}_{\rm NN(NNN)}$ is very strong, the valid range of $\Delta_3$ is relatively flexible. According to this analysis, the detunings that satisfy all the encoding constraints are chosen as $\Delta^{\rm max}_{1,2,3}/2\pi=2.5,10,15~{\rm MHz}$. Following the annealing ramp we obtain optimal graph colorings as an equal superposition of 10 degenerate optimal colorings with fidelity $70.4~\%$ as shown in Fig.~\ref{fig:K4_results}(d). Besides, there are two additional sets of 10 degenerate optimal solutions with the collection fidelity $16.9~\%$ and $7.8 ~\%$ whose configurations feature green and yellow atom at the center of the wheel, respectively. Hence, the total fidelity of the optimal graph colorings is up to $95.1\%$.

\section{Outlook and discussion} \label{mainsection:outlook}

In this paper, we have proposed qudit-based Rydberg atom arrays as a route to solve natively embedded MVGCPs. We employ a vertex-to-atom mapping where each color corresponds to a different Rydberg state. In this case, Rydberg quantum wires \cite{Kim2022RydbergProblems, Pichler2018QuantumArrays} are not required. In our simulation we include the long-range interaction tails. Qudit positioning is provided by an optical tweezer array. Unit-disk graphs are spatially rearranged into their compact structure where the numbers of equidistant edges are maximized. Our main results are as follows:

\emph{Planar graph coloring}--- We have analyzed the graph coloring on two types of planar graphs, equidistant and non-equidistant, respectively. We show that these $\chi(G)\le3$ graphs can be robustly solved using our 3-Rydberg optimizer, see Sec.~\ref{section:equidistant-max-degree3} and ~\ref{section:equidistant-max-degree4}. Our results show that different orders of graph symmetry correspond to different  degeneracies in the graph coloring solutions. Hence as suggested by \cite{Bernaschi2024TheGlass,Laumann2012QuantumTransitions,Janke1997Three-dimensionalTechniques}, quantum annealing could benefit from symmetry to alleviate the closing of the energy gap between the ground and first excited states as the number of qudits increases. However, due to the limitation of classical computer power, further work is needed to benchmark to larger system size. In Sec.~\ref{section:complete-K4}, the graph colorings on non-equidistant unit-disk graphs, e.g. wheel graphs with six vertices ($W_6$) and complete graphs with four vertices ($K_4$), are performed. Our results show that due to the effect of the negative inter-Rydberg interactions $V^{(ij)}$, the annealing yields the optimal colorings with lower fidelity compared to the equidistant case. For this reason, we propose a 3D graph embedding method for the $K_4$ graph where the equidistant structure is recovered using a tetrahedron. In this case, we demonstrate the highest order of degeneracy that any $4$-chromatic graphs could ever achieve, $24\,(4!)$.

\emph{Experimental Approach}---The proposed implementation of qudit-based annealing is compatible with current neutral atom experiments. To perform simultaneous excitation of multiple Rydberg states, we consider the case of $k=3$ low-power seed lasers being locked with independent frequency control relative to a common reference cavity. These can be combined prior to a high-power fiber amplifier stage, enabling common global amplitude control, with Rabi frequencies set by adjusting the relative power of the seed lasers. This can be used in conjunction with beam shaping techniques to enable homogeneous beam delivery across the atom array \cite{Ebadi2021QuantumSimulator}. For readout of the array, each Rydberg state can be mapped back to independent hyperfine-ground states on timescales fast compared to Rydberg state lifetime using STIRAP \cite{leseleuc19a,chen24} combined with fast ground-state rotations \cite{levine22}. State-selective imaging is then possible using sequential non-destructive imaging of the atom array \cite{kwon17,nikolov23}, with states shelved in the lower-hyperfine manifold prior to readout \cite{graham23}.

\emph{Potential for coloring non-planar graphs}---
Our focus has been on planar unit-disk graphs.
However, it is possible to extend our technique to solve more general planar graphs. For instance, in the context of solving MVGCPs on non unit-disk planar graphs, one can transform such planar graphs to corresponding unit-disk graphs by utilizing a Rydberg quantum wire. This is implemented by placing a chain of auxiliary atoms to connect vertices separated by more than the blockade length \cite{Kim2022RydbergProblems, Pichler2018QuantumArrays, Pichler2018ComputationalDimensions}. This approach has also been discussed in \cite{Clark1990UnitGraphs} in the context of finding unit-disk chromatic number. Here, the results of graph coloring on cycle graphs shown in Sec.~ \ref{section:cycle-graph} can be leveraged to implement the Rydberg quantum wire to solve MVGCPs on more complex planar graphs. According to the four-color theorem which states that every planar graph is 4-colorable \cite{Appel1977EveryReducibility,Appel1977EveryDischarging}, it suffices for one to use a $k=4$-Rydberg optimizer to solve MVGCPs on every planar graph augmented with Rydberg quantum wires. However, the challenge of solving MVGCPs on non-planar graphs with chromatic number greater than $4$ remains. Technically speaking, non-planarity spontaneously induces non-equidistant structure, in which the vertex-to-atom mapping will not be the most effective graph embedding, as the quantum annealing would suffer from mapped Rydberg-atom graphs having the negative inter-Rydberg interactions as previously addressed. Hence, augmenting such non-planar graphs with Rydberg quantum wires becomes a more strategic graph embedding method. However, at the crossing of the Rydberg quantum wires one needs to be aware of the intra-Rydberg interaction tails. On the contrary, if one insists on employing vertex-to-atom mapping, Rydberg states with significantly smaller inter-Rydberg interactions need to be found. Alternatively, one can employ 3D graph embedding, instead of 2D, to enhance the equidistant structure of such non-planar graphs, leading to a better system of encoding non-planar graph coloring problems. In terms of experimental feasibility, solving MVGCPs on $k$-chromatic graphs with $k>4$, i.e. general non-planar graphs, becomes extremely demanding. Apart from maintaining coherent control of individual atoms interacting with many lasers or microwave fields, it is also unlikely that we can find a larger set of same-parity Rydberg states compatible with a limited feasible range to satisfy all the encoding conditions previously mentioned in Sec.~\ref{section:Rydberg-encoding}. We shall leave this challenge as an open question for future research.  

\acknowledgements
We thank Jiannis K. Pachos for useful discussions on the many-body simulation aspects of the qudit-based Rydberg Hamiltonian; Antoine Browaeys and Giulia Semeghini for fruitful discussions on the experimental realization of Rydberg-atom quantum simulators; Mark Saffman for discussions on the additivity of Rydberg interactions; and Supanut Thanasilp on suggestions of numerical simulations. TA would like to express deep gratitude to Her Royal Highness Princess Maha Chakri Sirindhorn and the Thai government for providing PhD funding for the first author of this work. JDP and AD acknowledge support from the EPSRC through the following grants EP/T005386/1, EP/T001062/1 respectively. CSA acknowledge financial support provided by the UKRI, EPSRC grant Reference No. EP/V030280/1 (“Quantum optics using Rydberg polaritons”). The data presented in this work are available at \cite{datasetdoi}.

\bibliographystyle{apsrev4-2}
\bibliography{references}

\begin{thebibliography}{107}
\bibitem [{dat()}]{datasetdoi}%
  \BibitemOpen
  \bibinfo {note} {{DOI to be added in proof}}\BibitemShut {NoStop}%
\end{thebibliography}
\appendix
\begin{table*}[ht!]
    \centering
    \begin{tabular}{ c c c c c c c}
     \hline\hline
     Graph & $v_1$ & $v_2$ & $v_3$ & $v_4$ & $v_5$ & $v_6$  \\ [0.5ex] 
     \hline
     A & $(0,0,0)$ & $(a/2,a\sqrt{3}/2,0)$ & $(a,0,0)$ & - & - & -  \\ 
     
     B & $(0,0,0)$ & $(0,a,0)$ & $(a,a,0)$ & $(a,0,0)$ & - & -  \\
     
     C & $(0,0,0)$ & $(a/2,a\sqrt{3}/2,0)$ & $(a,0,0)$ & $(a/2,-a\sqrt{3}/2,0)$ & - & -  \\
     
     D & $(0,0,0)$ & $(a/2,a\sqrt{3}/2,0)$ & $(a,0,0)$ & $(3a/2,a\sqrt{3}/2,0)$ & $(2a,0,0)$ & - \\
     
     E & $(-a,0,0)$ & $(-a/2,a\sqrt{3}/2,0)$ & $(0,0,0)$ & $(a/2,a\sqrt{3}/2,0)$ & $(a,0,0)$ & $(0,a\sqrt{3},0)$  \\ 
     
     F& $(-a,0,0)$ & $(-a/2,a\sqrt{3}/2,0)$ & $(0,0,0)$ & $(a/2,a\sqrt{3}/2,0)$ & $(a,0,0)$ & $(3a/2,a\sqrt{3}/2,0)$  \\  
     
     G & $(0,a,0)$ & $(a\sqrt{3}/2,-a/2,0)$ & $(-a\sqrt{3}/2,-a/2,0)$ & $(0,0,0)$ & - & -  \\ 
     
     H & $(0,0,0)$ & $(0,a,0)$ & $(a,a,0)$  & $(a,0,0)$ & - & -  \\ 
     
     I & $(0,0,0)$ & $(-a,0,a)$ & $(0,a,a)$ & $(-a,a,0)$ & - & -  \\ 
     
     J & $(0,0,0)$ & $(a{\rm sin}\theta,a{\rm cos}\theta,0)$ & $(a{\rm sin}2\theta,a{\rm cos}2\theta,0)$ & $(a{\rm sin}3\theta,a{\rm cos}3\theta,0)$ & $(a{\rm sin}4\theta,a{\rm cos}4\theta,0)$ & $(a{\rm sin}5\theta,a{\rm cos}5\theta,0)$  \\ 
     \hline
     \hline
    \end{tabular}
    \caption{The cartesian coordinates of the vertices of all the problem graphs $G(V,E)$, where $V= \{v_1,v_2,... \}$, considered in the work. Here, the length $a$ of each problem graph is indicated in the corresponding figure in the main text with actual values given in Tab.~\ref{table:lattice-spacing}. For the graph J, $\theta=72^o$ for the pentagon.}
    \label{table:graph-implementation}
\end{table*}

\begin{table*}[ht!]
    \centering
    \begin{tabular}{ c c c c c c c c c c c }
     \hline\hline
     Graph & A & B & C & D & E & F & G & H & I & J  \\ [0.5ex] 
     \hline
     2-Rydberg optimizer & 5.26 & 5.26 & 4.99 & 5.26 & 4.91 & 5.26 & - & - & - & -  \\ 
     3-Rydberg optimizer & 6.33 & 6.41 & 6.75 & 6.75 & 6.75 & 6.75 & 3.37 & 4.45 & 5.61 & 4.10 \\
     \hline
     \hline
    \end{tabular}
    \caption{The actual lengths of lattice spacing $a$ in the unit $\mu {\rm m}$ used in all problem graphs for the 2- and 3-Rydberg optimizers.}
    \label{table:lattice-spacing}
\end{table*}

\section{Annealing Parameters}\label{Appendix:parameter-choosing}

This section provides additional details on choice of parameters used in the paper. Table.~\ref{table:graph-implementation} includes all the cartesian coordinates of each problem graph considered in this work, and Table.~\ref{table:lattice-spacing} gives the actual lengths of lattice spacing (in the unit $\mu {\rm m}$) used in the numerical simulations. 

\subsection{Parameter Choice}

The approach outlined above can be summarized as follows. First, having chosen a suitable set of Rydberg states with $\vert C_6^{(ij)}\vert <\vert C_6^{(i)}\vert,\vert C_6^{(j)}\vert$, the Rabi frequencies are chosen in the experimentally accessible regime of $1-10$~MHz such that the intra-Rydberg blockade lengths $R^{(i)}_b$ are comparable. Equidistant graphs can then be embedded using $R^{(ij)}_b < a <0.8R^{(1)}_b$ to ensure the blockade condition is met for nearest neighboring (NN) atoms with $|V^{(ij)}_{\rm NN}| \ll \Delta_i \ll V^{(i)}_{\rm NN}$, whilst experiencing a strong suppression of longer range next-nearest (NNN) and next-next-nearest (NNNN) neighbors $|V^{(ij)}_{\rm NNN(NNNN)}|,V^{(i)}_{\rm NNN(NNNN)} \ll \Delta_i$.

We consider as an example the 3-Fan (graph D), where the relative interaction strengths are given by $V_{\rm NN}^x=C_6^x/a^6$, $V_{\rm NNN}^x=C_6^x/(\sqrt{3}a)^6=V^x_{\rm NN}/27$ and $V_{\rm NNNN}^x=C_6^x/(2a)^6=V^x_{\rm NN}/64$ meaning these terms are strongly suppressed. Above we optimize above using $\Omega^{\rm max}_{1,2}/2\pi=3,7~{\rm MHz}$ with $R_b^{(1),(2),(12)}=7.02, 7.05$ and $5.15~\mu$m. This means we require $5.15<a<5.62~\mu$m, with $11.5\le V^{(1)}_{\rm NN}/2\pi\le19.1$~MHz, $27.5\le V^{(2)}_{\rm NN}/2\pi\le45.8$~MHz and $-5.0\le V^{(12)}_{\rm NN}/2\pi\le-3.0$~MHz. This introduces bounds on the detunings as $6.7\le\Delta^{\rm max}_1/2\pi\le11.5$~MHz and $6.7\le\Delta^{\rm max}_2/2\pi\le27.5$~MHz. Above we use $a=5.26~\mu$m and $\Delta^{\rm max}_{1,2}/2\pi=8,19$~MHz for implementing graph annealing. Note however that the largest of inter-Rydberg coupling in this range $|V^{(12)}_{\rm NN}|/2\pi=5.0$~MHz remains close to $\Delta^{\rm max}_1$ which can cause non-optimal or even invalid solutions to be lower in energy as discussed in Appendix \ref{Appendix:negativity-c6}.

\subsection{Parameter Robustness}

\begin{figure*}[ht!]
    \centering
    \includegraphics[width=15.5cm]{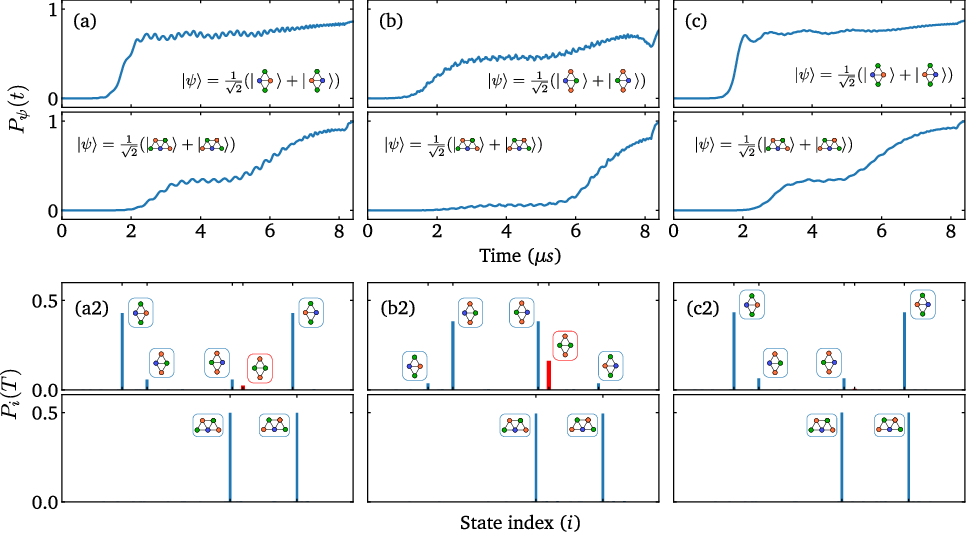}
    \caption{\textbf{Comparison of different driving parameters}. The three different driving protocols follow Eq.(\ref{eq:sweeping-protocol-detuning}-\ref{eq:sweeping-protocol-omega}) with (a) $\Delta^{\rm max}_1=\Delta^{\rm max}_2=10 \times 2\pi\,{\rm MHz}$, $\Omega^{\rm max}_1=\Omega^{\rm max}_2=3\times 2\pi \, {\rm MHz} $, (b) $\Delta^{\rm max}_1=\Delta^{\rm max}_2=10\times 2\pi \, {\rm MHz}$, $\Omega^{\rm max}_1=3\times 2\pi \, {\rm MHZ}$, $\Omega^{\rm max}_2=7\times 2\pi \, {\rm MHz}$ and (c) $\Delta^{\rm max}_1=8\times 2\pi \, {\rm MHz}$, $\Delta^{\rm max}_2=19\times 2\pi \, {\rm MHz}$, $\Omega^{\rm max}_1=\Omega^{\rm max}_2=3\times 2\pi \, {\rm MHz}$, have been performed to solve MVGCPs of the Diamond graph (C) and the 3-Fan graph (D). Here, the real-time annealing dynamics for the protocols (a) (b) and (c) have been simulated as Fig.(a), (b) and (c), respectively. The measurement results at the final time $t=T$ for the corresponding protocols follow in Fig.(a2), (b2) and (c2). Here, a variation in the energy of each Rydberg (color) state affects the ground configurations of solution states as happening in the Diamond graph (C). However, if the reflection symmetry of the graph requires a permutation between two atoms (vertices) excited in the two different energy-varying Rydberg states as happening in the 3-Fan graph (D), the ground configurations are robust to such changes in the driving parameters.}
   \label{fig:compare_driving}
\end{figure*}

To investigate the robustness of the annealing protocol to specific drive parameters, we show the 2-Rydberg optimization for the Diamond (C) and 3-fan (D) graphs as a function of drive parameters in Fig.~\ref{fig:compare_driving}. In (a) both states are driven with the same Rabi frequency and detuning with $\Omega^{\rm max}_{1,2}/2\pi=3~{\rm MHz}$, $\Delta^{\rm max}_{1,2}/2\pi=10$~MHz, whilst in (b) the Rabi frequency $\Omega^{\rm max}_{2}/2\pi=7$~MHz and finally (c) the optimum parameters $\Omega^{\rm max}_{1,2}/2\pi=3,7~{\rm MHz}$ and $\Delta^{\rm max}_{1,2}/2\pi=8,19$~MHz are used.

In each case as well as the temporal evolution, the final state decomposition is presented with valid solutions colored in blue and invalid colorings in red. Crucially, across this range of parameters the annealing protocol is robust in preparing valid MVGCP solutions with high fidelity, however analysis of the corresponding states shows that for the Diamond graph the use of equal detuning in (a) and (b) leads to weak population of an invalid coloring state featuring two greens at the centre of the Diamond due to the negative $C_6^{(12)}$. 

For the equal driving case (a) with $\Omega_1=\Omega_2$ and $\Delta_1=\Delta_2$, the preference for ground-state is defined by the next-nearest neighbor Rydberg interactions $V^{(i)}_{\rm NN} = C^{(i)}_6/(\sqrt{3}a)^6$. Since $C^{(1)}_6 < C^{(2)}_6$, this makes the green state a true ground state with an energy gap of $(C^{(2)}_6 - C^{(1)}_6)/(\sqrt{3}a)^6 = 1.2\times 2\pi \, {\rm MHz}$ from the state with green and orange inverted. By changing to $\Omega_2>\Omega_1$ in (b), it is possible to preferentially excite the orange state $\ket{r_2}$ earlier due to an effective suppression of the blockade radius with the increased Rabi frequency and a bias towards population of the state as the annealing profile crosses resonance. Finally in (c) with optimized choice of parameters giving $\Omega_2>\Omega_1$ and $\Delta_2>\Delta_1$ we recover good energy separation of the instantaneous eigenstates that prevents population of the invalid coloring states.

These results show that using the parameter choices above ensures optimal states are prepared, but the parameters are robust over small changes or fluctuations as required for experimental implementation.

\section{All about symmetry}\label{Appendix:symmetry}

\subsection{Symmetry-protected graph coloring solutions}
Since the Diamond (C) and 3-Fan (D) graphs possess four vertices, it is rational for one to expect the $\mathbb{S}_4$ symmetry in the graph coloring solutions. However, both of them are not complete graphs, according to the results shown in Fig.~\ref{fig:Planar_graph_and_driving}(f) and (g) they exhibit the $\mathbb{Z}_2$ symmetry caused by the reflection across the vertical axis of the graphs. In other words, there are only two permutations that commute with the Rydberg Hamiltonian in this case. Here, we state that the robustness of solutions to MVGCPs against variation in driving parameters, happens if these permutations involve exchanges between Rydberg-excited vertices. We confirm these arguments with the results shown in Fig.~\ref{fig:compare_driving}. Since the reflection of the 3-Fan graph (D) allows the permutations between vertices 1, 5 and vertices 2, 4, i.e. the set $\mathbb{Z}_2 =\{(), (15)(24) \}$, which are exactly the exchanges between the two Rydberg-excited vertices. Hence, the annealers in the three different protocols yield exactly the same coloring solutions as shown in the bottom panels of Fig.~\ref{fig:compare_driving}(a2), (b2) and (c2). In contrast, the reflection of the Diamond graph (C) allows the permutation between vertices 1, 3, i.e. the set $\mathbb{Z}_2 =\{(), (13) \}$, which is the exchange between the ground and Rydberg-excited vertices, such that the variations in the driving parameters result in the change in order of the energy of each Rydberg state. Hence, the graph coloring solutions to the MVGCP on the Diamond graph (C), as shown in the top panels of Fig.~\ref{fig:compare_driving}(a2),(b2) and (c2), are very specific to different driving parameters, and different solutions are therefore obtained in the three different protocols. 

\subsection{Influence on the energy spectrum of the triangular lattices}

In the analysis of the triangular lattice graphs in Fig.~\ref{fig:S3_results} we demonstrated that for the triangular graph with $\mathbb{S}_3$ symmetry we are able to color using either 2-Rydberg or 3-Rydberg optimization protocols, whilst for the ladder-shape graph only the 3-Rydberg optimizer yields valid results. This effect can be further seen as an artifact of symmetry protection.

\begin{figure}[t!]
    \centering
    \includegraphics[width=8.5cm]{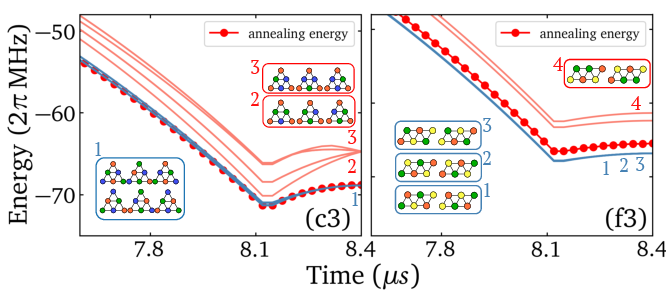}
    \caption{\textbf{Energy spectrum of the two triangular lattices.} As a continued result from Fig.~\ref{fig:S3_results}, the analysis of the energy spectrum in which the energy levels of the first twelve low-lying instantaneous eigenstates in the duration of $7.6-8.4 \, \mu s$ are plotted with respect to the following three cases: (c3) for the triangle-shape with the 2-Rydberg optimizer, and (f3) for the ladder-shape with the 3-Rydberg optimizers, respectively. Here, the sets of optimal degenerate graph colorings are labeled with blue color. while the sets of non optimal or incorrect graph colorings are labeled with red color. The number labeled in each group of degenerate states represent its energy order from lower to higher energy.}
    \label{fig:spectrum-analysis}
\end{figure}

To explore this we consider the low-lying instantaneous eigenstates for both cases during the final 7.6 to 8.4~$\mu$s of the annealing profile. For the triangular lattice in Fig.~\ref{fig:spectrum-analysis}(c3), the ground-state (blue) consists of six degenerate set of states with $\mathbb{S}_3$ symmetry, which remain well separated from the the excited state manifold throughout the sweep implying the symmetry is conserved $[H_{\rm Ryd},\mathbb{S}_3]=0$, whilst in contrast the low-lying excited states consist of invalid coloring states with neighboring $\ket{g}$ or $\ket{r_1}$ atom pairs with $\mathbb{Z}_3$ symmetry of three $2\pi/3$ rotations, which are separated during evolution but merge towards a denegerate manifold as the Rabi frequency ramps off. The red dotted line shows the annealing energy $\langle \Psi_a(t)|H_{\rm Ryd}(t)| \Psi_a(t)\rangle$ which clearly follows the ground-state $\mathbb{S}_3$ symmetry states.

For the ladder configuration in Fig.~\ref{fig:spectrum-analysis}(f3) we find $\mathbb{Z}_2$ ground states labeled by 1, 2 and 3 which are nearly degenerate despite the different ordering due to the rapid decrease of the NNN and NNNN Rydberg interactions in this graph configuration. The energy gap between the ground and the first excited state is large and remains consistent as the Rabi frequency is turned off, which is expected to benefit the quantum annealing for the system at larger sizes. 

\section{Influence of the negativity of the inter-Rydberg interaction $V^{(ij)}$}\label{Appendix:negativity-c6}

In the context of non-planar Rydberg-atom graphs in which connections to beyond-nearest-neighbor atoms are required, the connectivity leads to embedding as non-equidistant geometries such that there will be a significant increase in the magnitude of the negative nearest neighbor inter-Rydberg interaction $V^{(ij)}_{\rm NN} = C^{(ij)}_6/a^6$ with $C^{(ij)}_6<0$, which strongly affects the energy spectrum of the entire system. Since the feasible range of lattice spacing is subject to the encoding constraints, one would expect that it is challenging to extend the effective range of the Rydberg blockade to the next-nearest neighboring (NNN) atoms without violating the same conditions for the nearest neighboring (NN) atoms, unless one can find a pair of ideal (same-parity) Rydberg states $\ket{r_i}$ and $\ket{r_j}$ with a sufficiently small $C^{(ij)}_6$, ideally zero. 

Here, we provide an example of this scenario arising for MVGCP on the square $K_4$ graph as illustrated in Fig.~\ref{fig:K4_results}(b). We note that this graph is actually a planar graph, as it can be drawn without crossings. However, the crossing of next-nearest neighbor (NNN) interaction edges is essential for constructing larger non-planar graphs.
    
In this configuration, the 3-Rydberg optimizer with the same sweeping protocols described in Sec.~\ref{mainsection:equidistant-graph} is performed. The NN distances between nearest neighboring atoms are chosen as $a=4.45~\mu$m, which naturally yield the NNN distances between next-nearest neighboring atoms $\sqrt{2}a=6.29~\mu$m. At this spacing, the next-nearest neighbor (NNN) interactions are yielded as the following: $V^{(1),(2),(3)}_{\rm NNN}/2\pi=5.8,13.8,31.2$~MHz, and $V^{(12),(13),(23)}_{\rm NNN}/2\pi = -1.5,-0.6,-3.6$~MHz. With $\Delta_{1,2,3}^\mathrm{max}/2\pi=5,10,15$~MHz, the constraint in Eq.(\ref{eq:blockade-condition}) is satisfied for next-nearest neighboring (NNN) atoms. However, for the nearest neighboring (NN) atoms, there are strong inter-Rydberg interactions $V^{(12),(13),(23)}_{\rm NN}/2\pi = -12.0,-4.4,-28.5$~MHz which violate the encoding constraints. 

\begin{figure}[t!]
    \centering
    \includegraphics[width=8.0cm]{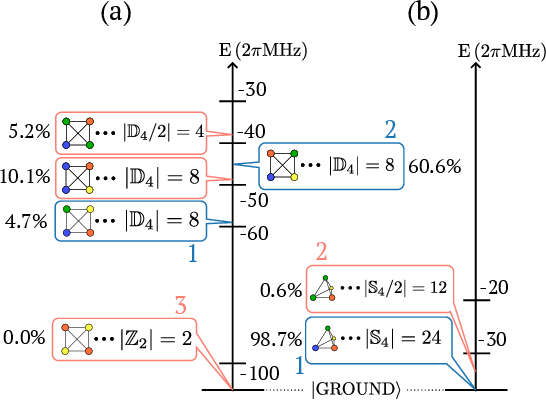}
    \caption{\textbf{Annealing collection at the final time of the $K_4$ graphs.} Given the significant impact of the negative inter-Rydberg interactions on the systems' energy spectrum, in the case of the square $K_4$ graph (a) the optimal graph colorings are obtained as the excited states denoted by the states 1 and 2 in the blue circles with total fidelity $65.3\%$, while the true ground state is denoted by the state 3 in the red circle. This negative effect can be suppressed in the 3D graph embedding as illustrated in (b) where the equidistant structure of the $K_4$ graph can be arranged into the tetrahedral graph. In this case, the twenty-four degenerate optimal graph colorings are obtained as the system's true ground state with fidelity $98.7\%$, as indicated by the state 1 in the blue circle.}
    \label{fig:K4_spectrum}
\end{figure}

The resulting energy spectrum at the end of the annealing ramp is shown in Fig.~\ref{fig:K4_spectrum}(a), along with the corresponding percentage population in the final annealing state, with valid solutions in blue boxes and invalid in red. In this case the dominant inter-Rydberg state interaction of $V^{(23)}_{\rm NN}/2\pi = -28.5$~MHz now causes the true ground-state to be the invalid case of orange and yellow on each corner, however the shift is so strong that it is not possible to populate this state during the annealing ramp as the nearest neighbor $\ket{r_2r_3}$ pair state is blockaded with $R^{(23)}_b > a$. Instead we see that the annealing profile preferentially prepares states with $\mathbb{D}_4$ symmetry of order 8, denoted by state 2 in the blue circle in Fig.~\ref{fig:K4_spectrum}(a), with fidelity up to $60.6\%$, whilst the alternative lower-lying valid $\mathbb{D}_4$ state (numbered 1) is also strongly suppressed by the $R^{(23)}_b > a$ blockade.

More generally, for graphs requiring strong NNN interactions, this inter-Rydberg blockade effect means even at $t=0$ there exist quantum states with lower-energy configurations than the trivial atomic ground state $\ket{gg...}$. For example, at initial time $t=0$ the actual ground energy of the $\mathbb{Z}_2$ states, denoted by state 3 in the red circle in Fig.~\ref{fig:K4_spectrum}(a), amounts to $-2\Delta_2-2\Delta_3 + 4 V^{(23)}_{\rm NN} + V^{(2)}_{\rm NNN} + V^{(3)}_{\rm NNN} = -19.0$ ($\times 2 \pi \, {\rm MHz}$). This leads to two problems: 1.) the adiabatic quantum annealing performed by starting the annealing from this state $\ket{gg...}$ can only adiabatically follow certain instantaneous excited states but not the true ground state of the system, 2.) the true ground state of the system in this case no longer encodes the solutions to the MVGCP on the square $K_4$ graph, as the solutions are now supposed to be lying in certain excited states of the Rydberg Hamiltonian. Despite the fact that our annealing algorithms have solved for a certain set of optimal graph coloring solutions, the annealing in this case breaks the conceptual definition of adiabatic quantum annealing. To amend this, one either need to find new actual Rydberg states such that the inter-Rydberg interactions $\vert V^{(ij)}\vert \ll1$ ( ideally zero) to ensure that the ground states of the new Rydberg Hamiltonian encodes the solutions to our interested MVGCP, or employ the 3D graph embedding to enhance the equidistant structure out of the $K_4$ graph, in which the twenty-four $\mathbb{S}_4$ degenerate optimal solutions, as the true ground state of the system, could occupy the annealing state at final time with fidelity $98.7\%$, as shown in Fig.~\ref{fig:K4_results}(c) and Fig.~\ref{fig:K4_spectrum}(b) without any disruption in the energy spectrum caused by the inter-Rydberg interactions.

\end{document}